\documentclass[aps,prc,twocolumn,superscriptaddress,floatfix,amsmath,amssymb,nofootinbib,longbibliography]{revtex4-2}

\usepackage{bm}
\usepackage{dcolumn}
\usepackage{amsfonts}
\usepackage{amsmath,amssymb}
\usepackage{yfonts}
\usepackage{mathrsfs}
\usepackage{amsthm}
\usepackage{bbold}
\usepackage{color}
\usepackage{braket}
\usepackage{hyperref}
\usepackage{graphicx}
\usepackage[caption=false]{subfig}

\usepackage{orcidlink}

\allowdisplaybreaks

\newcommand{\Xm}[1]{\ensuremath{#1_\text{max}}}

\newcommand{\magicint}{1.8/2.0 (EM)~}
\newcommand{\deltago}{$\Delta$NNLO$_\text{GO}$ (394)~}
\newcommand{\emarthuis}{1.8/2.0 (EM7.5)~}

\newcommand{\nnlosat}{NNLO$_\text{sat}$~}
\newcommand{\opt}{NNLO$_\text{opt}$~}
\newcommand{\texas}{N$^3$LO$_{\,\text{Texas}}$~}

\newcommand{\sixj}[6]{%
\begingroup
\setlength{\arraycolsep}{2pt} 
\renewcommand{\arraystretch}{1.2} 
\begin{Bmatrix}
#1 & #2 & #3\\
#4 & #5 & #6
\end{Bmatrix}
\endgroup}

\begin{document}

\title{Revisiting the Equation-of-Motion Method:\\
A Universal Framework for Correlated Quantum Systems}

\author{A.~Porro~\orcidlink{0000-0001-9828-546X}}
\email{andrea.porro@cea.fr}
\affiliation{Technische Universit\"at Darmstadt, Department of Physics, 64289 Darmstadt, Germany}
\affiliation{ExtreMe Matter Institute EMMI, GSI Helmholtzzentrum f\"ur Schwerionenforschung GmbH, 64291 Darmstadt, Germany}
\affiliation{CEA, DAM, DIF, 91297 Arpajon, France}
\affiliation{Université Paris-Saclay, CEA, Laboratoire Matière en Conditions Extrêmes, 91680, Bruyères-le-Châtel, France}

\begin{abstract}
A general implementation of the equation-of-motion (EOM) formalism for correlated many-body states is presented and applied to the description of collective excitations in atomic nuclei. While EOM approaches are traditionally formulated on top of independent-particle reference states, the present work extends the method to correlated reference states generated by modern many-body solvers. This formulation enables a consistent treatment of ground-state correlations and excited-state dynamics within a unified framework. Particular emphasis is placed on collective nuclear excitations employing chiral nuclear Hamiltonians in an ab initio context. The approach is motivated by the renewed interest in EOM techniques across several fields, including quantum chemistry and quantum computing, where they provide efficient and systematically improvable descriptions of excitation spectra. The present results demonstrate that the EOM framework offers a flexible and powerful tool for the microscopic description of nuclear spectroscopy beyond the traditional mean-field paradigm.
\end{abstract}

\maketitle

\section{Introduction}

The description of excited states remains one of the central challenges in quantum many-body physics. Across disciplines ranging from condensed-matter physics to quantum chemistry and nuclear physics, the theoretical description of excitations has become a cornerstone for understanding collective phenomena, symmetry breaking, and emergent degrees of freedom~\cite{Fetter03a,Negele94a,Ring1980_ManyBodyBook,Giuliani05a,Sachdev11a,Gonzalez20a,CoelloPerez24a}. In finite quantum systems in particular, excitation spectra encode detailed information about underlying interactions and many-body correlations, often providing stringent benchmarks for microscopic theories.

In nuclear physics, spectroscopy has historically played a decisive role in the development of theoretical frameworks for the atomic nucleus. Modern ab initio approaches, rooted in systematic nuclear interactions and many-body methods, have achieved remarkable success in the description of ground-state observables across broad regions of the nuclear chart~\cite{Hergert2020FP_AbInitioReview}. At the same time, the description of excited states remains a major challenge, especially for collective modes that involve coherent superpositions of many-particle configurations. The availability of chiral effective field theory Hamiltonians~\cite{Epelbaum09a,Machleidt11a,Hammer20a,Krebs20a,Miyagi2023EPJA_NuHamil} and systematically improvable many-body techniques has opened new opportunities for a unified microscopic description of nuclear structure and excitations within an ab initio framework.

Among the various theoretical approaches developed to describe excited states, the equation-of-motion (EOM) method occupies a particularly prominent place. Originally formulated in the context of nuclear theory~\cite{Rowe68a}, the EOM method is usually described introducing the concept of excitation operators acting on a reference state. Variants of the method have led to major developments in several fields, including Green’s-function approaches~\cite{McWeeny89a,Schirmer18a}, random-phase approximations (RPA)~\cite{Rowe68a}, and coupled-cluster (CC) theories~\cite{Stanton93a,Shavitt09a}. In nuclear physics, EOM ideas are traditionally associated with the mean-field plus (Q)RPA picture, where small-amplitude oscillations around an independent-(quasi)particle reference capture collective excitations with notable success~\cite{Ring1980_ManyBodyBook,Blaizot86a,Rowe2010a,Suhonen07a,Bacca26a}.

More recently, EOM methods have experienced a renewed interest beyond nuclear physics. In quantum chemistry, EOM-CC approaches have become among the most accurate and versatile tools for the description of electronic excited states, ionization spectra, and response properties, see Ref.~\cite{Musial20a} for a recent review. At the same time, renewed attention has emerged in the context of quantum computing, where the quantum-EOM has been proposed as an efficient strategy to extract excitation spectra from correlated variational quantum states~\cite{Ollitrault20a,Ollitrault21a,Motta22a,Motta24a,Hlatshwayo22a,Hlatshwayo23a}. These developments highlight the flexibility and broad conceptual relevance of the EOM framework for correlated quantum systems.

In this work, a general implementation of the EOM formalism for correlated many-body states is developed. In contrast to the standard formulation based on an uncorrelated mean-field reference, the present approach is designed to operate directly on correlated reference states generated by modern many-body solvers. This provides a unified framework in which correlations already incorporated in the ground state can be consistently propagated to the description of excited states. Particular emphasis is placed on collective excitations in nuclei, employing chiral nuclear Hamiltonians as microscopic input.

The objective is twofold. On the formal side, a general EOM framework applicable to correlated wave functions is established, without relying on the traditional independent-particle picture. On the physics side, the emergence and characterization of collective nuclear states from chiral interactions is investigated within an ab initio setting. By bridging concepts traditionally associated with many-body response theory and modern correlated wave-function approaches, the present work aims to contribute to the ongoing effort toward a predictive microscopic theory of nuclear excitations.

This paper is structured as follows: The theoretical basis of the EOM is presented in Sec.~\ref{sec:theo}, where the necessary truncations used later in this work are also specified. Section~\ref{sec:comparison} delivers a critical comparison to other many-body methods, underling similarities and differences with techniques used in ab initio nuclear theory. Numerical results are then shown in Sec.~\ref{sec:results}, focusing on the electric dipole response of oxygen isotopes. Eventually, a summary of the present development is given in Sec.~\ref{sec:conclusion}, where opportunities and future perspectives made possible by this work are also discussed. 

\section{Theoretical basis}
\label{sec:theo}
In this work many-body methods are classified according to which formulation of the Schr\"odinger equation they aim at solving. The solution of the \textit{time-independent} (TI) Schr\"odinger equation
\begin{equation}
\label{eq:sch_tind}
    H\ket{\Psi_\nu}=E_\nu\ket{\Psi_\nu}\,,
\end{equation}
delivers primarily ground-state properties, as well as low-lying stationary excited states. In Eq.~\eqref{eq:sch_tind} $H$ is the Hamiltonian of the system, $\{\ket{\Psi_\nu}\}_\nu$ the set of its eigenstates, and $E_\nu$ the corresponding eigen-energies. A wide class of methods attempting an approximate solution $\ket{\tilde{\Psi}}$ of Eq.~\eqref{eq:sch_tind} are founded on the variational principle, i.e., they require the stationarity of the energy functional
\begin{equation}
\label{eq:var1}
    \delta E[\tilde{\Psi}]= \delta\frac{\braket{\tilde{\Psi}|H|\tilde{\Psi}}}{\braket{\tilde{\Psi}|\tilde{\Psi}}}=0
\end{equation}
with respect to a family of variations of $\ket{\tilde{\Psi}}$. If Eq.~\eqref{eq:var1} holds for all possible variations $\ket{\delta\tilde{\Psi}}$, then $\ket{\tilde{\Psi}}$ is a solution of Eq.~\eqref{eq:sch_tind} and $E$ is an eigenvalue of $H$, otherwise it is an upper bound.

The \textit{time-dependent} (TD) Schr\"odinger equation
\begin{equation}
\label{eq:sch_tdep}
    i\partial_t\ket{\Psi(t)}=H\ket{\Psi(t)}
\end{equation}
delivers the system's dynamics and its response to external perturbations. A convenient and general starting point to develop a hierarchy of approximations converging to the full solution is given by Dirac's extremal condition of an action integral~\cite{Kramer81a,Kramer08a,Blaizot86a}
\begin{equation}
\label{eq:var2}
    \delta\int_{t_0}^{t_1}\left\langle\Psi(t)\left|i\partial_t-H\right|\Psi(t)\right\rangle dt=0\,.
\end{equation}
The contact point between the TI and the TD Schr\"odinger equation is given by stationary states of the form
\begin{equation}
    \ket{\Psi(t)}=e^{-iEt}\ket{\Psi(0)}\,,
\end{equation}
for which Eqs.~\eqref{eq:sch_tind} and~\eqref{eq:sch_tdep} coincide in the \textit{exact} limit. However, when the variation does not span the full Hilbert space, approximate solutions of Eqs.~\eqref{eq:var1} and~\eqref{eq:var2} will deliver different results.

The TD Eq.~\eqref{eq:var2} is chosen as a starting point of the present discussion. The exposition of the theoretical foundations of the EOM method and its derivation from the TD variational Eq.~\eqref{eq:var2} follow closely from Refs.~\cite{Rowe68b,Rowe80a}. While differing from more traditional derivations (see, for instance, Refs.~\cite{Rowe68a,Ring1980_ManyBodyBook,Suhonen07a}), the present discussion provides a deeper understanding of the geometrical foundations of the EOM method and highlights its differences from diagonalization techniques based on Eq.~\eqref{eq:var1}.

\subsection{Normal modes of the energy function}
Let now $\mathcal{H}$ be the energy function of the system on a submanifold of the full Hilbert space at some normalised initial state $\ket{\Psi}$
\begin{equation}
\label{eq:energy_funct}
    \mathcal{H}[\Psi]\equiv\braket{\Psi|H|\Psi}\,.
\end{equation}
Let us set, for convenience, $\ket{\Psi}\equiv\ket{\Psi(t=0)}$. Further constraints on the yet unknown $\ket{\Psi}$ will be discussed in Sec.~\ref{sec:approx}. 
The TD variational principle from Eq.~\eqref{eq:var2} reads, in terms of $\mathcal{H}$, as
\begin{equation}
\label{eq:eqdiff}
    d\mathcal{H}=i(\braket{d\Psi|\dot\Psi}-\braket{\dot\Psi|d\Psi})\,.
\end{equation}
Admitting only normalized trial functions, variations of $\ket{\Psi}$ can be formulated within the \textit{variational group} of unitary transformations, whose elements have the form 
\begin{equation}
\label{eq:unitary_group}
    U(t)=\exp{X(t)}\equiv \exp{x^a(t)X_a};\quad X^\dagger(t)=-X(t)\,,
\end{equation}
where Einstein's notation is used. The operators $X_a$ form an operator basis, while the coefficients $x^a(t)$ are TD scalar functions. Inserting Eq.~\eqref{eq:unitary_group} into Eq.~\eqref{eq:eqdiff} and expanding the energy function via the Baker–Campbell–Hausdorff formula yields, to linear order in $X$,
\begin{equation}
\label{eq:var4}
    \braket{\Psi|[X_a,H,X]|\Psi}-\braket{\Psi|[H,X_a]|\Psi}=\braket{\Psi|[X_a,i\partial_t X]|\Psi}\,.
\end{equation}
A step-by-step derivation of Eq.~\eqref{eq:var4} can be found in Appendix~\ref{app:group}. The symmetric form of the double commutator has been introduced as
\begin{equation}
    2[A,H,B]=[A,[H,B]]+[[A,H],B]\,.
\end{equation}
For a small-amplitude normal mode the anti-Hermitian operator $X(t)$ has the form
\begin{equation}
\label{eq:Xt}
    X(t)=\varepsilon(O^\dagger\, e^{-i\omega t}-O\, e^{i\omega t})\,,
\end{equation}
where $\varepsilon$ is a small parameter proportional to the amplitude of the oscillations. Equation~\eqref{eq:Xt} is the most general time-dependence of $X(t)$ for a TI Hamiltonian. Inserting Eq.~\eqref{eq:Xt} into Eq.~\eqref{eq:var4} and separating different orders of $\varepsilon$ and frequency components, the set of equations
\begin{subequations}
\label{eq:eom1}
    \begin{align}
    \braket{\Psi|[H,X_a]|\Psi}&=0\label{eq:stationary1}\\
        \braket{\Psi|[X_a,H,O^\dagger]|\Psi}&=\omega\braket{\Psi|[X_a,O^\dagger]|\Psi}\label{eq:eom_cc}\\
        -\braket{\Psi|[X_a,H,O]|\Psi}&=\omega\braket{\Psi|[X_a,O]|\Psi}
    \end{align}
\end{subequations}
is obtained, for all $X_a$. Equations~\eqref{eq:eom1} are the well-known equations of the EOM formalism~\cite{Rowe68a}. An alternative derivation not assuming the time-dependence from Eq.~\eqref{eq:Xt}, obtained by rewriting Eq.~\eqref{eq:eqdiff} in a canonical basis, is presented in Appendix~\ref{app:group}.

Equation~\eqref{eq:stationary1} is the Rayleigh-Ritz form of the variational principle from Eq.~\eqref{eq:var1} for the ground state and represents a stationary condition of the energy function
\begin{equation}
\label{eq:stationary}
    \frac{\partial\mathcal{H}}{\partial x^a}\Bigg|_{x=0}=0\,,\quad\quad\forall\,a\,.
\end{equation}
If the stationary condition in Eq.~\eqref{eq:stationary} is satisfied, Eqs.~\eqref{eq:eom1} yield the normal modes of the energy function $\mathcal{H}$, corresponding to small-amplitude periodic oscillations about a stationary point. It should be emphasized that Eqs.~\eqref{eq:eom1} (hence Eq.~\eqref{eq:var2}) admit solutions only when the metric is non-degenerate, i.e., on a symplectic manifold (see Appendix~\ref{app:symplectic}).

The excitation operators $O^\dagger$ are expanded on a generic many-body basis as
\begin{equation}
    O^\dagger_\nu=\sum_{a}[\mathcal{X}^a_\nu\eta^\dagger_a-\mathcal{Y}^a_\nu\eta_a]\,,
\end{equation}
where the $\nu$ index has been introduced to label different solutions, and the minus sign is chosen to keep a notation consistent with the usual formulation of the EOM method and RPA literature (see, for instance, Ref.~\cite{Ring1980_ManyBodyBook}). In such way, Eqs.~\eqref{eq:eom1} are written in the well-known form
\begin{equation}
\label{eq:eom_matrix}
    \begin{pmatrix}
        A & B \\
        B^* & A^*
    \end{pmatrix}
    \begin{pmatrix}
        \mathcal{X}_\nu\\
        \mathcal{Y}_\nu
    \end{pmatrix}
    =\hbar\omega_\nu
    \begin{pmatrix}
        U & V \\
        -V^* & -U^*
    \end{pmatrix}
    \begin{pmatrix}
        \mathcal{X}_\nu\\
        \mathcal{Y}_\nu
    \end{pmatrix}
\end{equation}
where
\begin{subequations}
\label{eq:matel}
    \begin{align}
        A_{ab}&\equiv\braket{\Psi|[\eta_a,H,\eta^\dagger_b]|\Psi}\,,\label{eq:A_def}\\
        -B_{ab}&\equiv\braket{\Psi|[\eta_a,H,\eta_b]|\Psi}\,,\label{eq:B_def}\\
        U_{ab}&\equiv\braket{\Psi|[\eta_a,\eta^\dagger_b]|\Psi}\,,\\
        -V_{ab}&\equiv\braket{\Psi|[\eta_a,\eta_b]|\Psi}\,,
    \end{align}
\end{subequations}
or in a more compact way
\begin{equation}
\label{eq:gep_mat}
    \mathsf{H}\,\mathsf{X} = \mathsf{N}\,\mathsf{X}\;\mathsf{\Omega}\,,
\end{equation}
with obvious notation. The super-matrices $\mathsf{H}$ and $\mathsf{N}$ are Hermitian, and so the matrices $A$ and $U$, while the $B$ matrix is symmetric and the $V$ matrix is anti-symmetric. Considerations about closure relations, orthogonalization of the solutions and sum rules of Eq.~\eqref{eq:eom_matrix} are discussed extensively in Refs.~\cite{Rowe68a,Rowe2010a}. One should stress that the set $\{\eta_\alpha\}$ is a many-body basis spanning, in principle, the whole Hilbert space. Equation~\eqref{eq:eom_matrix} reduces to the RPA equation only when the excitation space is limited to one-particle-one-hole ($1p$–$1h$) excitations and $\ket{\Psi}$ is the Hartree–Fock (HF) state. Otherwise, in the exact limit, it is a generalized eigenvalue problem that yields the exact normal modes of the system about the stationary state $\ket{\Psi}$. 
The solution of Eq.~\eqref{eq:eom_matrix} can be obtained with standard techniques also employed, for instance, in the Projected Generator Coordinate Method (PGCM)~\cite{Lathouwers76,Bally24a}. 

The central novelty of the present work is the fully general construction of the commutator operators entering Eqs.~\eqref{eq:matel} in a complete one-body operator basis.
This formulation is independent of the chosen initial state and can therefore be applied to any $\ket{\Psi}$. As a result, the equations of motion retain a universal operator structure valid for arbitrary reference states.

A related strategy has, to the author's best knowledge, mainly appeared in the context of quantum computing approaches~\cite{Ollitrault20a,Hlatshwayo22a,Hlatshwayo23a}, where operator dynamics are expressed in a basis-independent form for algorithmic implementations.

\subsection{Approximations}
\label{sec:approx}
Until now, no approximation has been introduced, so that Eq.~\eqref{eq:eom_matrix} exactly yields the normal modes of the energy function $\mathcal{H}$. In practice, however, two sources of approximation enter the solution of Eq.~\eqref{eq:eom_matrix}:
\begin{enumerate}
    \item the use of an initial state $\ket{\Psi}$ that is \textit{not} the exact ground state;
    \item the truncation of the operator set $\{\eta_a\}$.
\end{enumerate}

Regarding the first point, the EOM method does not uniquely define a strategy to determine the optimal state $\ket{\Psi}$ entering Eq.~\eqref{eq:eom_matrix}. The relevant question is therefore whether the chosen $\ket{\Psi}$ is a suitable reference with respect to the excitations spanned by the variational group, i.e., whether $\ket{\Psi}$ is stable under all considered variations. In other words, the initial state and the operator manifold must be chosen consistently: the correlations encoded in $\ket{\Psi}$ should correspond with the class of excitations represented by $X_a$.

If this is the case, $\ket{\Psi}$ is not only a stationary point but a minimum, and the Hessian $\mathsf{H}$ of the energy function $\mathcal{H}$ at $\ket{\Psi}$ is positive definite (or positive semi-definite in the presence of spurious zero-energy modes associated with symmetries~\cite{Rowe2010a}). Otherwise, if negative eigenvalues occur, the reference state is unstable and more closely resembles an excited state than the ground state. The connection between negative Hessian eigenvalues and phase transitions in finite nuclei is discussed in Refs.~\cite{Thouless61a,Rowe68b}.

Rowe~\cite{Rowe68a} suggested that the ground and excited states of the theory can be systematically improved by replacing $\ket{\Psi}$ with a new vacuum state $\ket{\Psi'}$, defined with respect to the set of de-excitation operators as
\begin{equation}
    O_\nu \ket{\Psi'} = 0 \,, \quad \forall \nu \, .
\end{equation}
The iterative solution of this condition, based on a generalized Thouless theorem for bosons~\cite{Ring1980_ManyBodyBook}, underlies the self-consistent RPA~\cite{Schuck20a}. No implementation of such method exists so far in realistic calculations, but the operatorial formulation of Eqs.~\eqref{eq:matel} represents a first step towards full self-consistency, suggesting further studies in this direction.

The present work focuses on the impact of correlations in the reference state on the response function. Results obtained with a mean-field reference state and a correlated reference state are compared. The former corresponds to the HF approximation, while the latter is constructed using the In-Medium Similarity Renormalization Group (IMSRG) approach~\cite{Tsukiyama10a,Hergert15a,Stroberg19a}, chosen as a representative correlation scheme.

For the excited-state basis, the full one-body operator space is considered, corresponding to the variational group of one-particle transformations (see Appendix~\ref{app:operator}). This choice is also numerically convenient, since the operators in Eqs.~\eqref{eq:A_def} and~\eqref{eq:B_def} remain two-body operators if the effective Hamiltonian is two-body.

Because the particle-hole distinction is not well defined for correlated states $\ket{\Psi}$, the full one-body operator space is used without restriction to the particle-hole ($ph$) sector.

\section{Comparison to existing methods}
\label{sec:comparison}
In the present EOM formulation, the commutators entering the matrix elements from Eqs.~\eqref{eq:matel} are defined and manipulated at the operator level. The excitation operator basis is taken as the full set of one-body operators and the expectation values of Eqs.~\eqref{eq:matel} are evaluated on a correlated IMSRG(2) reference state. The approach is fully general and allows for the propagation of ground-state correlations to the excited states.

It is instructive to compare this approach, together with the approximations discussed in Sec.~\ref{sec:approx}, to existing methods for quantum spectroscopy. Following the perspective outlined in Sec.~\ref{sec:theo}, the most natural classification is based on the variational equation one seeks to solve, namely the TI Schr\"odinger equation, Eq.~\eqref{eq:sch_tind}, or the TD Schr\"odinger equation, Eq.~\eqref{eq:sch_tdep}, and, more specifically, its normal modes. The discussion below is not intended to provide an exhaustive survey, but rather to offer a representative overview of methods commonly employed in nuclear theory, in particular within ab initio approaches. Time-dependent methods based on the direct integration of the TD Schr\"odinger equation are not considered here (see, e.g., Ref.~\cite{Bonaiti26a} for recent developments within the CC approach), nor are multi-reference approaches~\cite{Porro24a,Porro24b,Porro24c,Porro24d}.

Before proceeding, it is useful to emphasize the main message of this section. The TI variational principle of Eq.~\eqref{eq:var1} naturally gives rise to configuration-interaction-like equations and their associated approximations. In contrast, the TD variational principle of Eq.~\eqref{eq:var2} leads to RPA-like equations governing the normal modes of the system. Their most general form is given by Eq.~\eqref{eq:eom_matrix}, in which the Lie-algebra commutator metric emerges naturally. Furthermore, for a given reduced Hilbert subspace, the TD formulation involves a number of equations twice as large as in the TI case, since the dynamical character of the problem requires the introduction of a pair of canonically conjugate variables.

The distinction is perhaps most clearly illustrated by comparing the Tamm-Dancoff approximation (TDA) with the RPA~\cite{Rowe2010a,Ring1980_ManyBodyBook,Suhonen07a}. Both methods effectively explore the same Hilbert subspace, namely the space of $1p$-$1h$ excitations. However, whereas the TDA amounts to diagonalizing the Hamiltonian within this subspace, the RPA determines its normal modes, thereby providing a consistent treatment of the associated dynamical degrees of freedom.

\subsection{Configuration interaction}
\label{sec:CI}

The conceptually simplest approach to quantum spectroscopy is provided by the configuration interaction (CI) method. In this framework, the variational problem of Eq.~\eqref{eq:var1} is solved within a space of Slater determinants $\{\ket{\Phi_i}\}_i$,
\begin{equation}
    \ket{\Psi} = \sum_i c_i \ket{\Phi_i} \, ,
\end{equation}
where the states $\ket{\Phi_i}$ correspond to many-particle-many-hole excitations built on top of a reference state, typically the HF solution. Restricting the expansion to $1p$-$1h$ configurations yields the TDA~\cite{Ring1980_ManyBodyBook}, while the inclusion of higher-order $np$-$nh$ excitations leads to a hierarchy of increasingly sophisticated approximations.

In an $A$-body system, a diagonalization in the complete $Ap$-$Ah$ space yields the exact solution of the many-body problem. This strategy underlies the full configuration interaction (full-CI) approach and, in nuclear physics, the no-core shell model (NCSM)~\cite{Barrett13a}. If the diagonalization is instead performed within a restricted valence space (VS) of active particles, using either phenomenological interactions or effective Hamiltonians derived, for example, within the VS-IMSRG framework~\cite{Tsukiyama12a} (briefly discussed in Sec.~\ref{sec:eom-cc-imsrg}), one recovers the traditional shell-model (SM) approach~\cite{Caurier04a}.

Already at this stage, a general feature of CI-based methods becomes apparent: either the full Hilbert space is explored within a restricted active space, or a larger Hilbert space is considered while retaining only a selected subset of configurations, such as $1p$-$1h$, $2p$-$2h$, and so forth.

\subsection{EOM and VS extensions of CC and IMSRG}
\label{sec:eom-cc-imsrg}
Approaches closely related to the CI philosophy are employed within both CC and IMSRG frameworks for excited-state calculations. In CC theory, the many-body wave-function is written as
\begin{equation}
\ket{\Psi_{\text{CC}}}\equiv e^T\ket{\Phi}\,,
\end{equation}
where $T$ is a non-unitary cluster operator acting on a reference state $\ket{\Phi}$, typically the HF solution. This defines the similarity-transformed Hamiltonian
\begin{equation}
\bar{H}\equiv e^{-T}He^T.
\end{equation}
Since similarity transformations preserve eigenvalues, the EOM-CC method~\cite{Hagen10a,Hagen14a} and related approaches such as LIT-CC~\cite{Bacca14a,Marino25a} obtain excited states by diagonalizing $\bar{H}$ in a truncated CI space. In nuclear applications this is typically limited to $2p$-$2h$ configurations, with extensions to $3p$-$3h$ sectors~\cite{Miorelli18a}. Due to the large dimensionality, the diagonalization is usually performed iteratively, most often using Lanczos methods~\cite{Caurier04a}.

A similar strategy is adopted within the IMSRG~\cite{Tsukiyama10a,Hergert15a,Stroberg19a}. In the Magnus formulation, the correlated ground state is
\begin{equation}
\ket{\Psi_{\text{IMSRG}}}
\equiv
\lim_{s\to\infty}U^\dagger(s)\ket{\Phi}
=
\lim_{s\to\infty}e^{\Omega(s)}\ket{\Phi}\,,
\end{equation}
where $U(s)$ is a unitary flow that decouples the Hamiltonian. In EOM-IMSRG~\cite{Parzuchowski16a,Parzuchowski17a}, excited states are obtained by diagonalizing
\begin{equation}
H(s)=U(s)HU^\dagger(s)
\end{equation}
in a truncated CI space. Both EOM-CC and EOM-IMSRG are systematically improvable, as enlarging the truncation leads to convergence to the exact solution.

The CI character of EOM-CC was already noted in its original formulation~\cite{Stanton93a} and is well documented~\cite{Shavitt09a}. The CI structure arises from the restriction of the excitation operator space: while a complete basis would include both excitation ($ph$) and de-excitation ($hp$) operators, only excitation operators are retained~\cite{Stanton93a}. Since these commute with $T$, the problem reduces to a CI-like eigenvalue equation.

This omission effectively removes half of the canonical degrees of freedom required for a fully dynamical description, reducing the formulation to a CI-like structure. While equivalence holds in the full Hilbert space, truncated realizations differ in their treatment of collective dynamics.

The restriction to excitation operators was formally justified within unitary CC theory by Mukherjee and collaborators~\cite{Prasad85a,Datta93a} (see also Refs.~\cite{Asthana23a,Kim23a,Reinholdt24a}), where a so-called self-consistent basis allows a CI-like reformulation of the EOM problem.

Physically, this reflects the equivalence of TI [Eq.~\eqref{eq:sch_tind}] and TD [Eq.~\eqref{eq:sch_tdep}] formulations for stationary states in the exact limit. In practice, however, this equivalence is lost at the approximate level, and EOM-CC is structurally closer to Eq.~\eqref{eq:sch_tind} than to Eq.~\eqref{eq:sch_tdep}. It would therefore be interesting to investigate to what extent the truncation of the Hilbert space and the choice between TI and TD formulations commute, and how the resulting EOM problem differs between the traditional formulation of EOM-CC and -IMSRG and the one presented in this work.

Eventually, related CC and IMSRG approaches construct effective VS Hamiltonians by decoupling the VS and diagonalizing it exactly, in analogy with SM calculations. This underlies SM-CC and VS-IMSRG; a review is given in Ref.~\cite{Stroberg19a}.

\renewcommand{\arraystretch}{1.7}

\begin{table*}[]
\caption{Summary of the main features of different EOM-inspired theories, see the discussion is Sec.~\ref{sec:discussion}. The references point to first implementations of the presented methods within ab initio nuclear theory. Indices $p$ and $h$ run on particle and hole states respectively, while indices $a$ and $b$ run on the whole one-body basis.}
\label{tab:scheme}
\begin{ruledtabular}
\begin{tabular}{rclcc}
 & Hamiltonian & Variational Eq.& Reference state $\ket{\Psi}$ & Excitation operator basis \\
\hline
EOM-IMSRG~\cite{Parzuchowski16a} & $H(s)= U(s)HU^\dagger(s)$ & TI Eq.~\eqref{eq:var1} &$\ket{\Phi}$ &$\sum_{php'h'}X^{php'h'}c_{p'}^\dagger c_{h'}c_p^\dagger c_h\ket{\Phi}$ \\
EOM-CC~\cite{Hagen10a} & $\bar{H}= e^{-T}He^T$ & TI Eq.~\eqref{eq:var1} & $\ket{\Phi}$ & $\sum_{php'h'}X^{php'h'}c_{p'}^\dagger c_{h'}c_p^\dagger c_h\ket{\Phi}$\\
This work & $H$ & TD Eq.~\eqref{eq:var2} & $\ket{\Phi(s)}=U^\dagger(s)\ket{\Phi}$ & $\sum_{ab}X^{ab}c_a^\dagger c_b\ket{\Phi(s)}$ \\
\end{tabular}
\end{ruledtabular}
\end{table*}

\subsection{Green's function theory}
Finally, an important comparison can be drawn with Green's function theory. It can be shown (see Chapters 13 of Refs.~\cite{McWeeny89a,Schirmer18a}) that the exact polarization propagator is recovered as the solution of Eq.~\eqref{eq:eom_matrix} when the complete operator basis is used. This highlights that the RPA-like structure of Eq.~\eqref{eq:eom_matrix} is not an artifact of the present derivation, but arises naturally in the description of dynamical properties of many-body systems. Accordingly, Eq.~\eqref{eq:eom_matrix} provides a natural starting point for spectroscopic approximation schemes.

The connection with propagator theory is also useful for assessing the approximation level associated with the operator basis. Restricting to the full one-body space corresponds, from this viewpoint, to a generalized RPA, where the polarization propagator is built from resummed particle-hole bubble diagrams. The difference from standard RPA is that the underlying one-body propagators inherit the correlations encoded in the reference state $\ket{\Psi}$, a scheme referred to as dressed RPA.

For a HF reference state, the formalism reduces to standard RPA. More correlated choices of $\ket{\Psi}$, such as the IMSRG(2) wave-function used here, effectively dress the propagators and generate more advanced approximations. This idea has been explored in Ref.~\cite{Raimondi18a} in the context of Self-Consistent Green's Functions. However, a drawback of that work relates to the representability of the self-consistent propagator via an effective propagator with a reduced number of poles: different strategies for selecting these poles introduce a dependence of the final results on the chosen procedure. This ambiguity does not arise here, since the present formulation is based directly on the wave-function rather than on propagators, and leads to a well-defined eigenvalue problem without any explicit pole selection.

\subsection{Discussion}
\label{sec:discussion}
The choice made in the present work to restrict the operator basis to the Lie algebra of one-body operators may appear limiting. As argued in Ref.~\cite{Duguet22a}, consistency between ground- and excited-state descriptions for correlated states, as obtained in CC or IMSRG frameworks, would at least require the inclusion of $2p$-$2h$ excitations. This viewpoint is closely related to the EOM-CC and EOM-IMSRG approaches discussed in Sec.~\ref{sec:eom-cc-imsrg}. As previously noted, these methods are of CI type~\cite{Stanton93a,Shavitt09a}, where the original Hamiltonian is replaced by a similarity-transformed operator. In both cases, the ground state is given by a HF determinant, while excited states are expanded in $1p$-$1h$ and $2p$-$2h$ configurations, as illustrated in Tab.~\ref{tab:scheme}.

Independently of the group-theoretical considerations of Appendix~\ref{app:group}, and of whether a TI or TD formulation is adopted, the key difference with the present approach lies in the treatment of correlations. In EOM-CC/EOM-IMSRG, excited states are built on an uncorrelated HF reference, while ground-state correlations are encoded implicitly in a similarity-transformed Hamiltonian. In contrast, the present framework constructs excited states directly on a correlated reference state.

As a result, ground-state correlations are transferred to excited states in a non-perturbative and basis-independent way. This reflects that the EOM formulation addresses the response problem, i.e., how a state $\ket{\Psi}$ reacts to a given perturbation, rather than the eigenvalue problem for $H$. When the variational space is identified with operators $\eta$, the response is determined by the corresponding class of variations, which can be chosen on physical grounds. In particular, the Lie algebra of one-body transformations generates the entire space of accessible one-body density variations. Since it includes local operators and their conjugate momenta, that are known to efficiently capture collective dynamics~\cite{Reinhard90,Reinhard92,Reinhard92a}, one expects strongly collective states to be accurately described within this space, provided their treatment is consistent.

Finally, a complementary numerical perspective can be given. EOM-CC and EOM-IMSRG rely on approximate diagonalizations in large configuration spaces (typically up to $3p$-$3h$), where the dominant cost lies in the diagonalization step. In the present work, the computational effort is shifted to the evaluation of matrix elements, particularly the construction of double-commutator operators and their expectation values on a correlated state. Once computed, the resulting small eigenvalue problem can be solved exactly at negligible cost, while preserving the symplectic structure of the theory.

\section{Results}
\label{sec:results}

\subsection{Numerical details}
The EOM problem~\eqref{eq:eom_matrix} is solved, in this work, within the one-body transformation group employing an IMSRG(2)~\cite{Tsukiyama10a,Hergert15a,Stroberg19a} reference state $\ket{\Psi}$. In order to evaluate the matrix elements entering Eq.~\eqref{eq:eom_matrix} in a completely general fashion, the single and double commutators need to be developed in an operatorial form. The detailed derivation in an angular-momentum coupled framework is given in Appendix~\ref{app:Jscheme}, and it follows previous developments from Refs.~\cite{Chen93a,Chen93b,LuJohnson2018,Porro25a,Porro26a}. The correct implementation of the operators was benchmarked by a comparison with the RPA matrices $A$ and $B$, element wise. Indeed, when the expectation value is taken with respect to the uncorrelated HF reference state, the RPA is recovered. The benchmark was performed for several multipolarities.

Once the double-commutator is built, the expectation value is taken with respect to a correlated reference state (IMSRG(2) in this work). Eventually, after all the matrix elements have been evaluated (which is trivially parallelizable), the generalized eigenvalue problem from Eq.~\eqref{eq:gep_mat} is easily solved by using the \texttt{python}
function \texttt{numpy.eig}.

In the following, all displayed calculations are performed within a harmonic oscillator basis, with a frequency of $\hbar\omega=16$~MeV. The convergence is gauged upon variation of the number of major shells included, i.e., \Xm{e}$=(2n+l)_{\text{max}}$. The same \Xm{e} truncation is used for the reference state and for the excitation operator basis. Three-nucleon interactions are included using the normal-ordered two-body approximation~\cite{Hergert15a,Heinz2020,Frosini21d}. To ensure convergence, the number of configurations is allowed up to $e_1+e_2+e_3\leq E^{(3)}_{\text{max}}=24$, by exploiting the developments of Ref.~\cite{Miyagi2022PRC_NO2B}. The NN and 3N matrix elements are computed using the \texttt{NuHamil} code~\cite{Miyagi2023EPJA_NuHamil}, while IMSRG(2) calculations were performed employing the \texttt{imsrg++} code~\cite{Stroberg2024_IMSRGGit}.

The following sections focus on the postive branch of the isovector dipole response 
\begin{equation}
    R(Q_{1}^{\text{IV}},\omega)\equiv\sum_{\mu\nu}\frac{|\braket{\Psi_\nu|Q_{1\mu}^{\text{IV}}|\Psi_0}|^2}{\omega-\omega_\nu} \label{eq:response}
\end{equation}
in $^4$He and $^{16}$O, for which the centre-of-mass corrected transition operator is employed
\begin{equation}
\label{eq:dipole_int}
    Q^\text{IV}_{1\mu}=\frac{N}{A}\sum_{i=1}^Zr_iY_{1\mu}(\hat{r}_i)-\frac{Z}{A}\sum_{i=1}^Nr_iY_{1\mu}(\hat{r}_i)\,.
\end{equation}
In Eq.~\eqref{eq:dipole_int} ${r}_i$ is the radial position of the $i$-th particle, while $Y_{\lambda\mu}(\hat{r}_i)$ are the spherical harmonics~\cite{Varshalovich88a} of rank $\lambda$; $Z$, $N$ and $A$ are, respectively, the number of protons, neutrons and nucleons in the system. In the following, the spectral function, obtained from Eq.~\eqref{eq:response} as
\begin{equation}
\label{eq:lorentz}
    S(Q_{1}^{\text{IV}},\omega)=-\frac{1}{\pi}\text{Im}\,R(Q_{1}^{\text{IV}},\omega)\,,
\end{equation}
is displayed with a small imaginary component added to the energy argument, i.e., the quantity $S(Q_{1}^{\text{IV}},\omega+i\Gamma)$ is shown, where $\Gamma$ is a smearing parameter (imaginary energy) whose value is set to 1~MeV unless stated otherwise.

\begin{figure}
    \centering
    \includegraphics[width=\columnwidth]{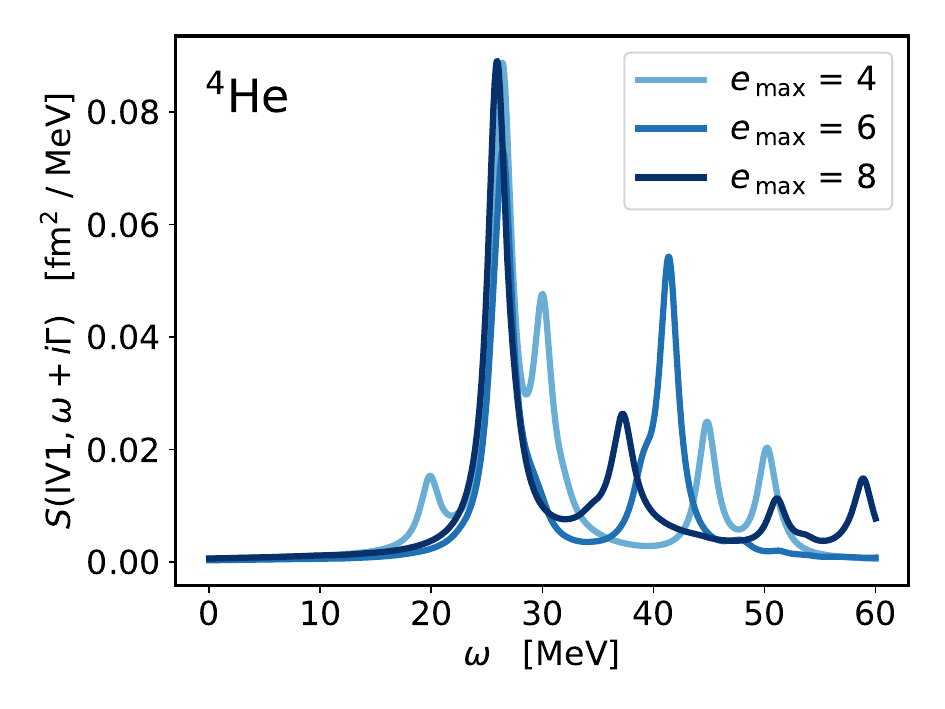}
    \caption{Isovector dipole response in $^{4}$He obtained employing the EOM technique based on an IMSRG(2) reference state and a full one-body operator basis for the excited states. Calculations were performed employing the \opt\cite{Ekstrom13a} interaction with a harmonic oscillator frequency of $\hbar\omega=16$~MeV. Different model space sizes are compared.}
    \label{fig:He4}
\end{figure}

\subsection{Benchmark calculations for $^4$He}

The dipole response of $^{4}$He is computed employing the \opt interaction~\cite{Ekstrom13a}. This allows to employ the following results as a benchmark, by comparison to NCSM calculations of the same quantity performed with the same interaction~\cite{Burrows23a}. Results from the present calculations are displayed in Fig.~\ref{fig:He4} for several model space sizes. A fast convergence of the giant dipole resonance position and height is observed. It locates at 26~MeV and it compares well with NCSM calculations, where it was reported at 26.5~MeV. 

To give a sense of the numerical effort, in an $\Xm{e}=8$ calculation, the number of basis elements in the $J^\pi=1^-$ channel is $N=380$ (one-body channels allowed by the selected symmetry), which translates in $\frac{1}{2}N(N+1)=72390$ matrix elements to be evaluated for every matrix $A$, $B$, $U$ and $V$. This number is much larger than the $ph$ basis employed in standard RPA calculations because the full one-body operator space is spanned (also including $hh'$ and $pp'$ configurations). However, the problem is embarrassingly parallelizable, such that with future optimization bigger model spaces will certainly be of easier access.

\subsection{Dipole response in oxygen isotopes}

\begin{figure}
    \centering
    \includegraphics[width=\columnwidth]{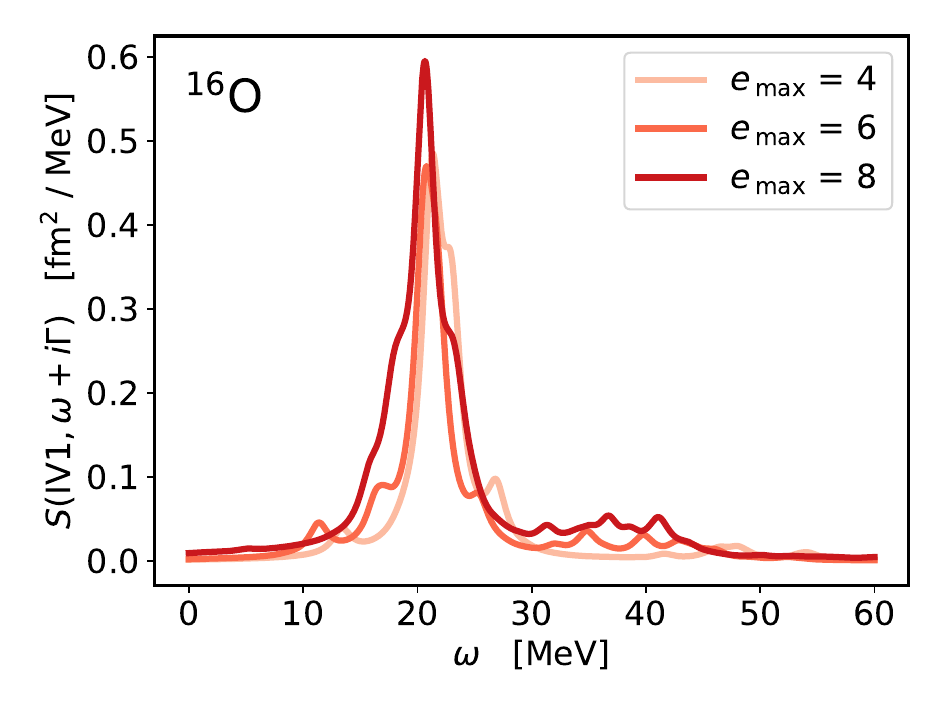}
    \caption{Isovector dipole response in $^{16}$O obtained employing the EOM technique based on an IMSRG(2) reference state and a full one-body operator basis for the excited states. Calculations were performed employing the \texas\cite{Hu25a} interaction with a harmonic oscillator frequency of $\hbar\omega=16$~MeV. Different model space sizes are compared.}
    \label{fig:O16_Texas}
\end{figure}

After assessing the reliability of the present method in $^4$He with a simplified NN interaction, the isovector dipole response in $^{16}$O is now addressed by employing several chiral NN plus 3N interactions. The convergence of the numerical results is gauged using the recently developed \texas interaction~\cite{Hu25a}. Figure~\ref{fig:O16_Texas} displays the isovector dipole response in $^{16}$O for several values of \Xm{e}. One can observe that the giant dipole resonance peak quickly stabilizes at about 21~MeV, even if a better convergence for smaller structures, especially in the range between 30 and 40~MeV would require a bigger model space size. The rapid convergence of the giant dipole resonance in this relatively small model space is aligned with recent observations in time-dependent CC calculations~\cite{Bonaiti26a}.

\begin{figure}
    \centering
    \includegraphics[width=\columnwidth]{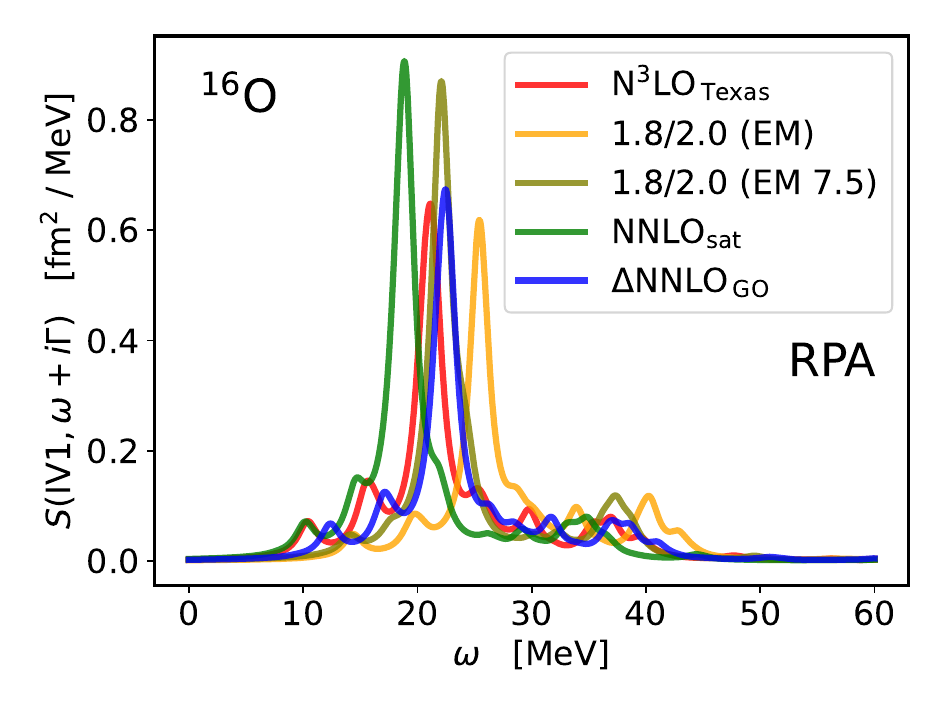}\\[-4mm]
    \includegraphics[width=\columnwidth]{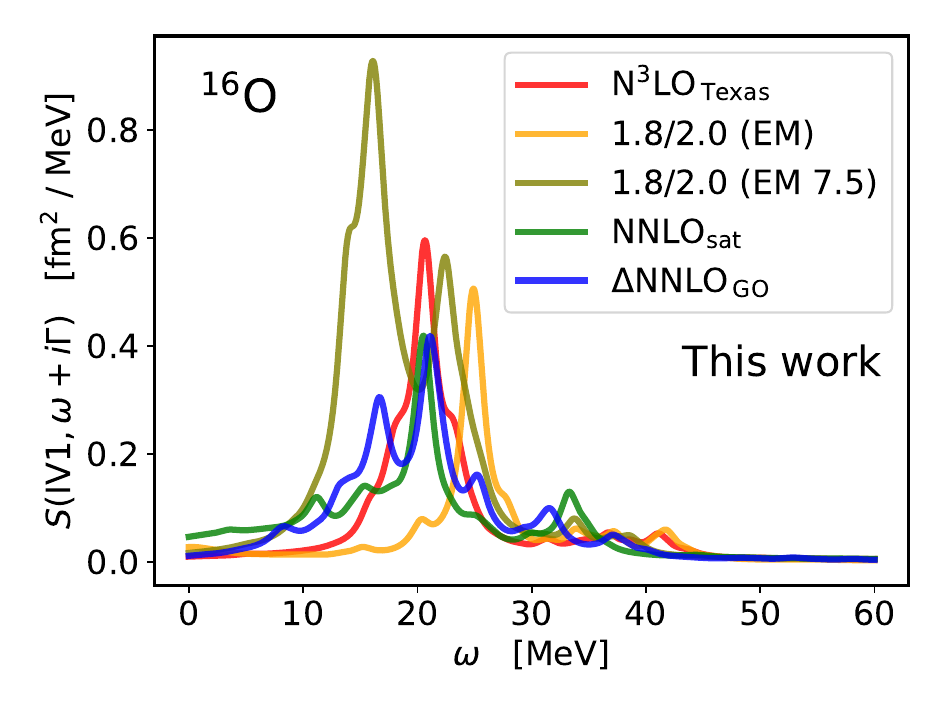}
    \caption{Isovector dipole response in $^{16}$O from RPA calculations (upper panel) and the EOM technique based on an IMSRG(2) reference state and a full one-body operator basis for the excited states (bottom panel). Calculations are performed at \Xm{e}=~8 and $\hbar\omega=16$~MeV. 
    }
    \label{fig:O16_EOM}
\end{figure}

\begin{figure}
    \centering
    \includegraphics[width=\columnwidth]{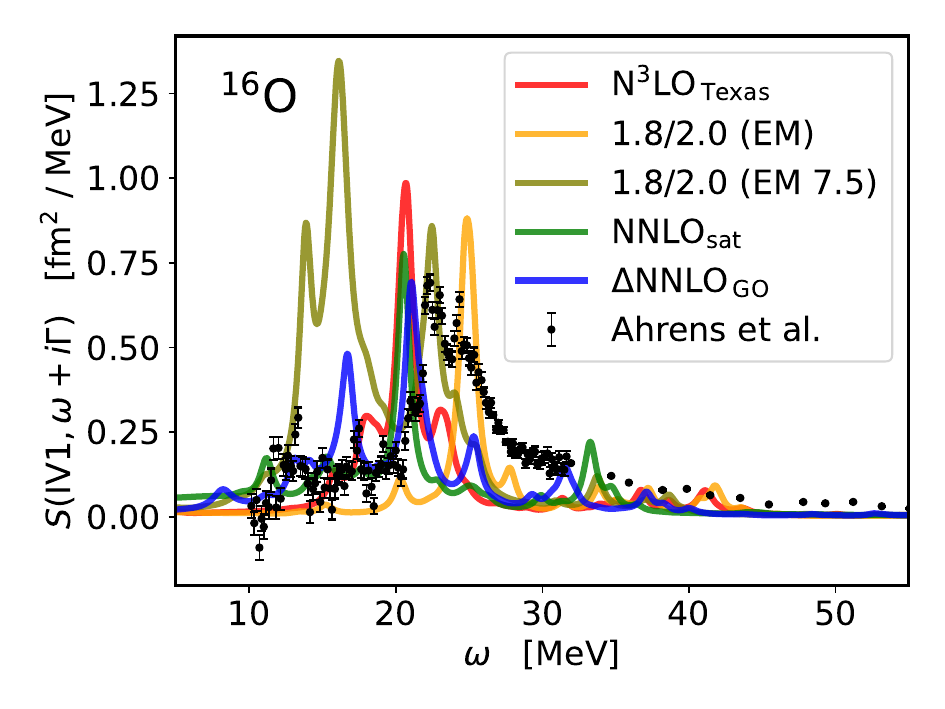}
    \caption{Same as the bottom panel of Fig.~\ref{fig:O16_EOM} but for $\Gamma=0.5$~MeV. Experimental data are taken from Ref.~\cite{Ahrens75a}.}
    \label{fig:O16_exp}
\end{figure}

Calculations are then repeated employing other chiral interactions, specifically \nnlosat\cite{Ekstrom13a}, \deltago\cite{Ekstrom15a}, \magicint\cite{Hebeler10a} and \emarthuis\cite{Arthuis24a}. As a starting point, RPA calculations are shown in the upper panel of Fig.~\ref{fig:O16_EOM} for comparison. One clearly observes a quite large dispersion of about 7~MeV relative to the position of the main peak, with \nnlosat providing the smallest value (about 18.5~MeV) and \magicint the largest (about 25.5~MeV). The same interactions are then probed employing the method developed in this work, whose results are shown in the bottom panel of Fig.~\ref{fig:O16_EOM}. The dispersion of the positioning of the main peak between different interactions is reduced to 4~MeV, and only 1.7~MeV if one excludes the \magicint results. Eventually calculations employing \nnlosat\cite{Ekstrom13a}, \deltago\cite{Ekstrom15a}, \emarthuis\cite{Arthuis24a} and \texas\cite{Hu25a} interactions provide a giant dipole resonance at about 21~MeV. It is interesting to observe that the inclusion of higher many-body correlations by means of the present framework strongly reduces the interaction dependence displayed at the mean field + RPA level, especially for the harder \nnlosat interaction. In some cases the correlated calculations also display a significative increase in the strength of lower-lying states. This effect is specially pronounced for calculations employing the \deltago and \emarthuis interactions. While the magnitude of such response may be exaggerated, it is true that there is experimental evidence of a very rich excitation landscape in the region between 10 and 20~MeV. This can be observed in Fig.~\ref{fig:O16_exp}, where the same calculations as Fig.~\ref{fig:O16_EOM} are reported with a different smearing parameter of $\Gamma=0.5$~MeV and are compared to experimental data from photoabsorption cross section~\cite{Ahrens75a}.

With the increased resolution and a direct comparison to experimental data one can appreciate the good performance of the \emarthuis~interaction, while the \magicint~overpredicts the giant dipole resonance energy by more than 2~MeV and the other three interactions (\texas, \nnlosat, \deltago) underpredict it by about 1.5~MeV. The nice behaviour of the \emarthuis can be correlated with its good performances in terms of neutron skin predictions~\cite{Arthuis24a}, which is well known phenomenologically to be closely related to the dipole response~\cite{Centelles08a,Piekarewicz12a}.

\begin{figure}
    \centering
    \includegraphics[width=\columnwidth]{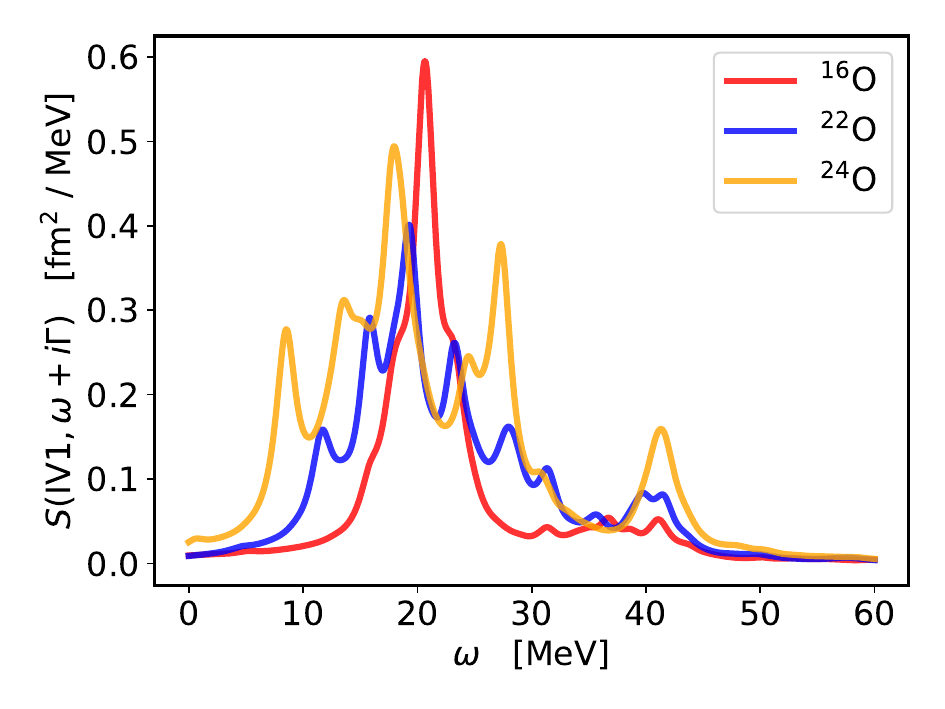}
    \caption{Isovector dipole response in closed-shell oxygen isotopes obtained employing the EOM technique based on an IMSRG(2) reference state and a full one-body operator basis for the excited states. Calculations were performed employing the \texas\cite{Hu25a} interaction with a harmonic oscillator frequency of $\hbar\omega=16$~MeV at \Xm{e}$=8$.}
    \label{fig:Oiso}
\end{figure}

Eventually the dipole response in other closed-shell oxygen isotopes is investigated in Fig.~\ref{fig:Oiso} employing the \texas interaction. Neutron-rich $^{22}$O and $^{24}$O display a shift of the giant dipole resonance (main peak) towards lower energies, correlated with the neutron excess. Moreover, several structures appear at lower energies and the response is generally more fragmented than in $^{16}$O. While more thorough convergence studies should be repeated as for $^{16}$O, the qualitative behaviour is coherent with the common knowledge about dipole response in neutron-rich nuclei, in absence of precise experimental photoabsorption data in such systems. 

\subsection{Complex solutions}

\begin{figure}
    \centering
    \includegraphics[width=\columnwidth]{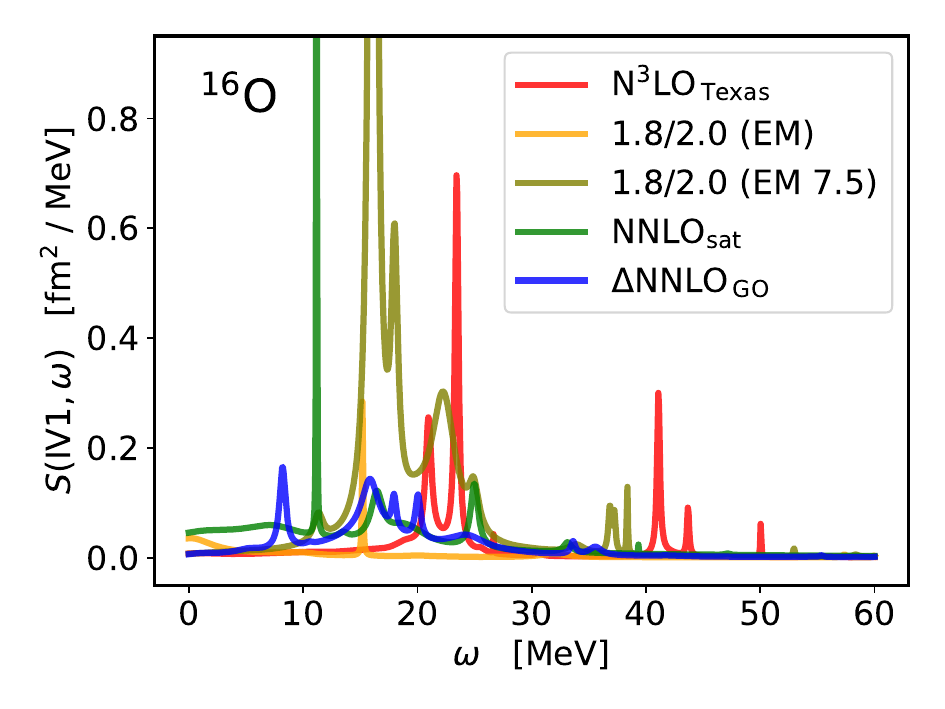}
    \caption{Complex part of the dipole response in $^{16}$O, as obtained by setting $\Gamma=0$~MeV. The calculation parameters and interactions are the same as in Fig.~\ref{fig:O16_EOM}.}
    \label{fig:O16_complex}
\end{figure}

\begin{figure}
    \centering
    \includegraphics[width=\columnwidth]{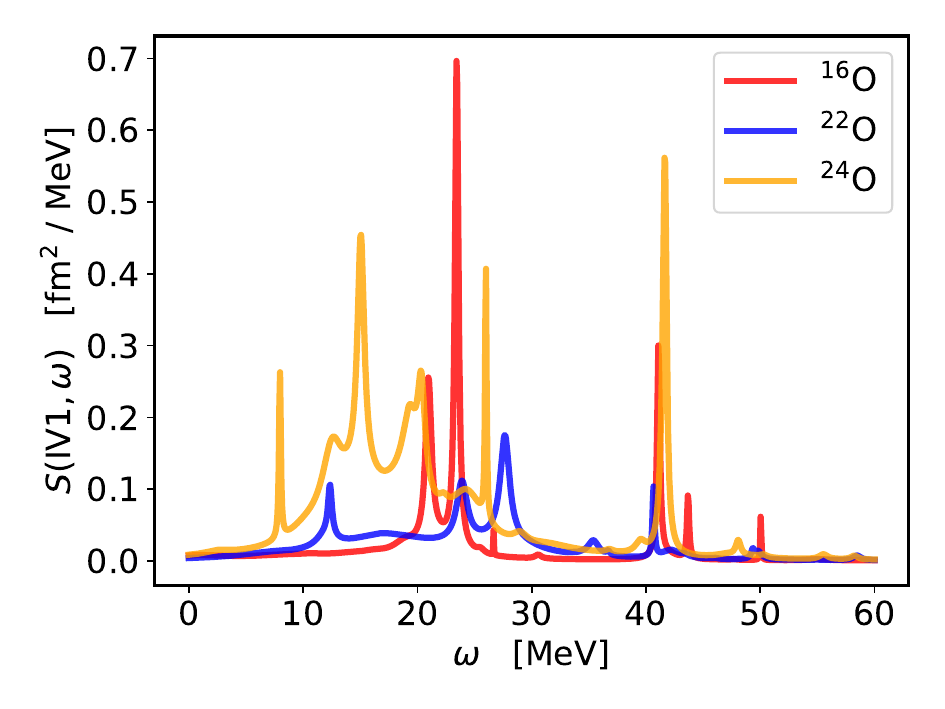}
    \caption{Complex part of the dipole response in closed-shell O isotopes, as obtained by setting $\Gamma=0$~MeV. The calculation parameters and interaction are the same as in Fig.~\ref{fig:Oiso}}
    \label{fig:Oiso_complex}
\end{figure}

When Eq.~\eqref{eq:eom_matrix} is solved using a correlated IMSRG(2) reference state, a significant number of complex-energy solutions is obtained, in addition to real ones. In standard RPA, complex eigenvalues are typically regarded as unphysical~\cite{Thouless60a,Thouless61a,Rowe68b}, as they signal an instability of the reference state with respect to the chosen class of excitations. For instance, in deformed nuclei described from a spherical HF reference, instabilities appear in the quadrupole sector, reflecting the fact that the spherical state is not stable against quadrupole deformations. More generally, similar instabilities can arise in other multipole channels, depending on the nucleus, as a manifestation of long-range correlations not captured at the mean-field level.

In the present case, complex eigenvalues are associated with one-body excitations beyond the $ph$ sector. Both the HF and IMSRG(2) reference states are stable with respect to $ph$ variations, but not with respect to general one-body transformations. In particular, excitations of the form $c_a^\dagger c_b$, with both $a$ and $b$ above (or both below) the HF Fermi surface, are not represented in either reference state\footnote{While a $0^+$ reference state is, by symmetry, expected to be stable against first-order variations with $J^\pi\neq0^+$, different $J^\pi$ sectors of the one-body algebra are not closed subalgebras (except for $0^+$). In particular, two non-$0^+$ excitations can couple to $0^+$, and through the Jacobi identity this affects the double-commutator structure, so that instabilities in the full one-body sector may manifest across all $J^\pi$.}. When restricted to $ph$ excitations, the Hessian is positive semi-definite and the spectrum is real; however, this subspace is incomplete and does not form a closed group. Such instabilities reflect the non-variational character of IMSRG and are also known in CC theory~\cite{Sherrill98a,Mizukami20a,Marie21a}.

The appearance of complex solutions in Eq.~\eqref{eq:eom_matrix} is therefore consistent with the general properties of RPA-like problems~\cite{Ullah71a,Barillier09a,DePace14a,Nakada16a,Nakada16b}. Although Hermitian operators have real spectra, Eq.~\eqref{eq:eom_matrix} yields normal modes of the energy functional in a non-orthogonal operator space, and complex eigenvalues arise when the Hessian matrix in Eq.~\eqref{eq:gep_mat} is not positive (semi) definite. The resulting spectrum typically contains real pairs, purely imaginary pairs, and complex quartets. One consequence is the breakdown of sum-rule exhaustion, even when the one-body basis is complete~\cite{Nakada16b}.

Addressing these instabilities requires incorporating missing correlations in the reference state. Possible routes include self-consistent RPA-like schemes~\cite{Schuck20a} and orbital-optimization approaches such as orbital-optimized CC~\cite{Sherrill98a,Mizukami20a,Marie21a}, where the energy is made stationary with respect to one-body rotations. This is equivalent to the strategy proposed in Ref.~\cite{Kutzelnigg92a} within stationary perturbation theory.

From this perspective, the emergence of complex solutions should not be interpreted as a pathology of Eq.~\eqref{eq:eom_matrix}, but rather as a diagnostic of an incomplete description of correlations in the reference state within the chosen operator space. In particular, it signals that the assumed state $\ket{\Psi}$ does not provide a stable extremum of the energy functional under the full set of one-body variations, even though it may remain stable within restricted subspaces such as the $ph$ sector. The complex branches therefore encode physically relevant information on missing correlation channels and on the reorganization of the ground state induced by collective degrees of freedom. In this sense, their appearance points directly to the need for an improved, symmetry-consistent optimization of the reference state rather than to a failure of the formalism itself.

For completeness, the complex part of the excited-states spectrum of calculations shown in previous sections is displayed here. While for real-energy solutions an artificial imaginary part was always needed for visualising the response function, according to Eq.~\eqref{eq:lorentz}, complex energies already encodes their own smearing, such that for complex transition energies one can set $\Gamma=0$~MeV. Complex solutions with small imaginary parts will then display a sharper peak, while stronger broadening is associated to energies with large imaginary components. 

The complex part of the dipole strength function in $^{16}$O is displayed in Fig.~\ref{fig:O16_complex} for the same calculation parameter and interactions as in Fig.~\ref{fig:O16_EOM}. One immediately observes very different behaviours, according to the employed interaction. The \magicint interaction, for instance, displays almost no strength in the complex sector, while a very large amount is observed for the \emarthuis interaction. Interestingly, most of the complex strength manifests in the same region as the giant dipole resonance in the real counterpart. This may suggest a coupling between the main resonance and one-body excitations above the Fermi surface.

Eventually, Fig.~\ref{fig:Oiso_complex} displays the complex counterpart of Fig.~\ref{fig:Oiso} for different (closed-shell) oxygen isotopes. In particular $^{24}$O displays a rich behaviour in the 10-20~MeV region, with several poles lying close to the real axis with a non-negligible transition strength, producing many resonances with different widths.

\section{Conclusions and perspectives}
\label{sec:conclusion}
In this work, a fully general implementation of the EOM method was presented. The commutators entering the master equation were constructed at the operator level within the complete variational group of one-body operators, independently of the chosen reference state. As a result, the equations of motion retain a universal operator structure valid for arbitrary reference states, making the formalism readily applicable to any correlated wave-function approach.

Following existing literature, the link between the EOM method and the TD variational principle has been reviewed in Sec.~\ref{sec:theo}, stressing its intrinsic dynamical nature. 
Analogies and differences with existing many-body methods allowing for an approximate determination of excitation spectra were discussed in Sec.~\ref{sec:comparison}, with particular emphasis on EOM-CC and -IMSRG. 
Calculations based on an IMSRG(2) reference state were performed and discussed in Sec.~\ref{sec:results}, displaying globally a reasonable agreement with available experimental data for $^{16}$O. More neutron-rich oxygen isotopes were also studied, displaying lower-lying strength. The newly developed framework allowed to appreciate the presence of one-body instabilities within the correlated (non-variational) IMSRG(2) state, encouraging further studies to variationally optimize the employed reference state in the space of explored excitations.

In this respect, the implementation of two-body operators (whose matrix elements are derived in Appendix~\ref{app:Jscheme}) allowing to compute the Hessian of the energy function in the space of one-body operators sets the ground for orbital-optimization techniques~\cite{Sherrill98a,Marie21a} within nuclear theory calculations. While having long been considered numerically intractable in the quantum chemistry community, new hybrid calculations in the quantum realm are reviving this technique~\cite{Mizukami20a}, and the present works shows that improvements in this direction are possible in ab initio nuclear theory calculations as well. Another possible and non-orthogonal strategy in the framework of ground-state optimization may follow from self-consistent RPA strategies~\cite{Schuck20a}, for which the availability of Hessian operators constitute a first essential step. 

More systematic calculations of other systems and of different excitation nature (including charge-exchange excitations) are of easy access, given the completely general implementation of the present technique. Further numerical optimization will allow to reach bigger model spaces (hence heavier system) at a milder computational cost. Another possible strategy to be explored relates to the group of scale transformations, which may be suited for studying very collective phenomena, like multi-phonon states. Additionally, a straightforward application of this formalism is the calculation of transition between excited states, which is of interest for de-excitation studies~\cite{Goriely18a,Goriely25a}.

The universal nature of the framework presented in this work suggests several avenues for future studies employing different reference states, including symmetry-breaking and projected mean-field states, which would provide the first realistic implementation of a projected (Q)RPA formalism~\cite{Federschmidt85a,Porro23a}. Even with symmetry-broken reference states the number of basis operators would stay tractable thank to the $J$-coupled representation of the matrix elements operators. Not least, this method could be easily transferred to modern techniques allowing for a hybrid quantum-classical computational treatment of the many-body ground state, like the quantum variational eigensolver~\cite{Ollitrault20a,Ollitrault21a,Carrasco26a,Hlatshwayo22a,Hlatshwayo23a}.

Additional fragmentation of the response function, allowing to take into account decays of the phononic states and the associated damping~\cite{Bortignon98a} (giant resonances lying above the particle-emission threshold) could be obtained by enlarging the operator basis, allowing to account for, e.g., two-body transformations. This strategy, however, would break the group-variational structure of the theory (while remaining likely intractable as for its computational cost), unless the full $A$-body transformations are explored. Similarly, particle-vibration approaches may be envisioned~\cite{Niu16a,Colo22a,Li23a,Lv26a}, but the computational burden is likely to be excessive. A more appealing possibility may derive from a statistical treatment of the coupling between phonons and decay channels, as done for instance in Ref.~\cite{Severyukhin18a,Severyukhin21a,Arsenyev23a}.

The EOM technique, as implemented in this work, provides a universal framework for computing spectral functions in many-body quantum systems, accessible to all methods capable of evaluating expectation values of two-body operators. While further numerical optimization remains to be explored, the approach is naturally parallelizable and offers a unique route to accessing the Hessian of the energy functional at arbitrary reference states, paving the way towards orbital optimization techniques in ab initio nuclear theory and highlighting the central role of one-body physics in strongly correlated systems.

\section*{Acknowledgements}
The author thanks A. Schwenk for valuable feedback and support throughout this work. Sincere thanks are also extended to G.~Col\`o, T. Duguet and V. Som\`a for valuable comments on the manuscript. Discussions with M. Drissi, T. Papenbrock, P.-G. Reinhard, X.~Roca-Maza, R. Roth and A.~Tichai are gratefully acknowledged. This work is supported by the Deutsche Forschungsgemeinschaft (DFG, German Research Foundation) - Project-ID 279384907 - SFB 1245. The author gratefully acknowledges the Gauss Centre for Supercomputing e.V. (www.gauss-centre.eu) for funding this project by providing computing time on the GCS Supercomputer JUWELS at J\"ulich Supercomputing Centre (JSC).

\appendix

\section{Quantum dynamics}
\label{app:symplectic}
The starting point of the EOM is the possibility of associating to every quantum system a \textit{quantum phase space}, i.e., a differentiable manifold with a suitable geometrical structure. This phase space is shown to be the projective Hilbert space $P\mathscr{H}$, endowed with the structure of a K\"ahler manifold~\cite{Kramer81a,Cirelli91a,Schroeck10a}.

A Hilbert space $\mathscr{H}$ is as a complex linear space with a Hermitian inner product $\braket{\Psi|\Phi}$. In order to treat it as a phase space, it is taken as a real linear space of twice the dimension, where every basis state $\ket{\psi_a}$ is replaced by the pair $\ket{\psi_a}$ and $i\ket{\psi_a}$. The realified Hilbert space has then two non-degenerate forms, $\text{Re}\braket{\Psi|\Phi}$ and $\text{Im}\braket{\Psi|\Phi}$. The former is symmetric and positive-definite, while the latter is anti-symmetric and closed, i.e., \textit{symplectic}, which makes $\mathscr{H}$ a phase space, in parallel to classical Hamilton dynamics.

Following Ref.~\cite{Rowe80a}, let now $\mathcal{H}$, as defined in Eq.~\eqref{eq:energy_funct}, be the energy function of the system on a submanifold $M$ of the full Hilbert space $\mathscr{H}$. The TD variational principle from Eq.~\eqref{eq:var2} reads, in terms of $\mathcal{H}$, as in Eq.~\eqref{eq:eqdiff}. On a general coordinate chart $(x^1,x^2,\ldots)$ for a neighbourhood of $M$, Eq.~\eqref{eq:eqdiff} reads then
\begin{align}
    \frac{\partial\mathcal{H}}{\partial x^a}&=-2\,\text{Im}\left\langle\frac{\partial\Psi}{\partial x^a}\bigg|\dot\Psi\right\rangle\nonumber\\
    &\equiv\sigma_{ab}\dot x_b\,,
\end{align}
where $\sigma$ is the symplectic metric
\begin{equation}
    \label{eq:metric}
    \sigma_{ab}=-2\,\text{Im}\left\langle\frac{\partial\Psi}{\partial x^a}\bigg|\frac{\partial\Psi}{\partial x^b}\right\rangle\,.
\end{equation}
If the metric is non-degenerate, $\sigma_{ab}$ can be inverted, so to give Hamilton's equations of motion
\begin{equation}
    \dot x^a=\sigma^{ab}\frac{\partial\mathcal{H}}{\partial x^b}\,,
\end{equation}
and $M$ is said to be symplectic. We conclude that \textit{the action integral in Eq.~\eqref{eq:var2} only has extremal paths on a symplectic submanifold}.

Pure states of a system are represented by unit vectors of $\mathscr{H}$, and two unit vectors $\ket{\Psi}$ and $\ket{\Phi}$ differing only by a phase factor
\begin{equation}
\label{eq:equivalence}
    \ket{\Psi}= e^{i\delta}\ket{\Phi}
\end{equation}
represent the same physical state. The projective Hilbert space $P\mathscr{H}$ is defined as the set of equivalence classes of unit vectors with respect to Eq.~\eqref{eq:equivalence} and is the quantum analogue of the phase space of classical mechanics. By passing to the projective Hilbert space $P\mathscr{H}$, one precisely removes possible degeneracies from $\sigma$, such that the induced form becomes non-degenerate.

\section{Variational group and local charts}
\label{app:group}
Once the symplectic structure associated with Eq.~\eqref{eq:var2} has been established, the variational space must be specified, i.e., a local chart for $P\mathscr{H}$ must be introduced in the neighbourhood of a given state $\ket{\Psi}$. A formal discussion is provided in Ref.~\cite{Rowe80a}, which develops the group-theoretical foundations underlying the following statements.

We shall admit only normalized trial functions. Variations of $\ket{\Psi}$ that conserve the norm can be formulated as unitary transformations, and one can require that such transformations form a group, which is referred to as the \textit{variational group} $\mathscr{G}$. The group $\mathscr{G}$ is unitary, with an associated Lie algebra $\mathscr{L}$. For $\ket{\Psi}$ \textit{any} state in $\mathscr{H}$, the orbit of $\mathscr{G}$ in $\mathscr{H}$ containing $\ket{\Psi}$ is the set
\begin{equation}
    M=\{\ket{\Psi(g)}=g\ket{\Psi};\,g\in \mathscr{G}\}\,,
\end{equation}
and any $g\in\mathscr{G}$ can be written as
\begin{equation}
\label{eq:group_var}
    g=\exp{X};\quad X^\dagger=-X;\quad X\in\mathscr{L}_r\subset\mathscr{L}_c\,,
\end{equation}
where $\mathscr{L}_r$ and $\mathscr{L}_c$ are, respectively, the real and complex Lie algebras of anti-Hermitian operators $X$, $\mathscr{L}_c$ being the full Lie algebra associated with $\mathscr{G}$.
Upon introduction of the equivalence relation~\eqref{eq:equivalence}, one can show that the orbit $M$ containing $\ket{\Psi}\in P\mathscr{H}$ is itself a submanifold of $P\mathscr{H}$.
Thus, a coordinate chart $(x^a)$ for a neighbourhood of $\ket{\Psi}\in M$ is defined by
\begin{equation}
\label{eq:chart}
    \ket{\Psi(x)}\equiv\exp{(x^a X_a)}\ket{\Psi}\,,
\end{equation}
where Einstein's notation is used and the $X_a$ are a basis of $\mathscr{L}_r$ or $\mathscr{L}_c$, equivalently~\cite{Kutzelnigg92a}. Eventually, with the local chart defined by Eq.~\eqref{eq:chart}, the symplectic metric~\eqref{eq:metric} at $\ket{\Psi}$ is given by
\begin{equation}
\label{eq:metrix_X}
    \sigma_{ab}=-i\braket{\Psi|[X_a,X_b]|\Psi}\,,
\end{equation}
while the energy function can be expanded by means of the Baker–Campbell–Hausdorff formula, delivering
\begin{equation}
    \mathcal{H}[\Psi+\delta\Psi]=\mathcal{H}[\Psi]+\delta\mathcal{H}[\Psi;\delta\Psi]+\frac{1}{2}\delta^2\mathcal{H}[\Psi;\delta\Psi]+O((\delta\Psi)^3)\,,
\end{equation}
where
\begin{subequations}
\label{eq:H_expansion}
    \begin{align}
        \delta\mathcal{H}[\Psi;\delta\Psi]&=\sum_a\braket{\Psi|[H,X_a]|\Psi}x^a\,,\\
        \delta^2\mathcal{H}[\Psi;\delta\Psi]&=\sum_{ab}\braket{\Psi|[X_a,H,X_b]|\Psi}x^a x^b\,.\label{eq:Hessian}
    \end{align}
\end{subequations}
Eventually, with the definitions from Eqs.~\eqref{eq:metrix_X} and~\eqref{eq:H_expansion} at hand, Eq.~\eqref{eq:eqdiff} reads, at linear order in $X$, as Eq.~\eqref{eq:var4}~\cite{Rowe68b}.

By bringing the Hessian of the energy function into canonical form, i.e., finding a basis $(X_a)=(iQ^a,-iP_a)$ such that
\begin{subequations}
\label{eq:canonical}
    \begin{align}
        \braket{\Psi|[Q^a,Q^b]|\Psi}&=\braket{\Psi|[P_a,P_b]|\Psi}=0\,,\\
        \braket{\Psi|[Q^a,P_b]|\Psi}&=i\delta^a_b\,,\\
        \braket{\Psi|[Q^a,H,Q^b]|\Psi}&=\delta_{ab}B^a\,,\\
        \braket{\Psi|[P_a,H,P_b]|\Psi}&=\delta^{ab}C_a\,,\\
        \braket{\Psi|[Q^a,H,P_b]|\Psi}&=0\,,
    \end{align}
\end{subequations}
at the stationary point, one obtains a set of equations fully equivalent to Eq.~\eqref{eq:eom1}. This follows from the equivalence of employing a complex or real basis of the variational group~\cite{Kutzelnigg92a}.

\section{Example of variational groups}
\label{app:operator}
The choice of the operator basis for the variational group is discussed in detail in Ref.~\cite{Kutzelnigg92a}. Because the exponential chart entering Eq.~\eqref{eq:chart} stems from group theoretical arguments, the basis elements $X_a$ of the variational group shall form a Lie algebra. Among many-body operators, this criterion greatly limits the space of possible choices. For an $N$-body system, among the available options one finds:
\begin{enumerate}
    \item one-particle transformations;
    \item $N$-particle excitations;
    \item scale transformations.
\end{enumerate}
Higher many-body transformations (but not $N$-body), like two-particle transformations of the form
\begin{equation}
    U=\exp{\left\{\sum_{abcd}X_{abdc}\,c_a^\dagger c_b^\dagger c_c c_d\right\}}\,,
\end{equation}
as used in the so-called second-RPA~\cite{Yannouleas86a,Takayanagi88a,Drozdz90a,Gambacurta16a,Gambacurta26a}, do not unfortunately constitute a unitary group (except, for the given example of two-body transformations, for two-particle systems), since the basis operators do not form a Lie algebra. While providing a natural extension to the group of one-body transformation, increasing the spectral fragmentation, the group property is lost.

Focusing now on the variational group of one-particle transformation, its generators, i.e., a basis of $\mathscr{L}_c$, are the one-particle excitation operators
\begin{equation}
\label{eq:Lie_basis}
    c_a^\dagger c_b\,.
\end{equation}
The indices $a$ and $b$ in Eq.~\eqref{eq:Lie_basis} run on the whole single-particle space, without distinction between particle and hole states.
The Lie group so generated is known as $U(n)$, the unitary group of dimension $n$.
The corresponding basis in $\mathscr{L}_r$ would be
\begin{equation}
    c_a^\dagger c_b-c_b^\dagger c_a\,,\quad i(c_a^\dagger c_b+c_b^\dagger c_a)\quad\text{for}\;a< b\,;\quad ic_a^\dagger c_a\,.
\end{equation}
A further extension (not considered in this work) of the Lie algebra of operators from Eq.~\eqref{eq:Lie_basis}, is given by the inclusion of
\begin{equation}
\label{eq:pp_qq}
    c_a^\dagger c_b^\dagger\,;\quad c_a c_b\,.
\end{equation}
The enlarged set still represents a Lie algebra, but the corresponding Lie group, the $SO(2n)$ group (special orthogonal group of dimension $2n$) is not particle-number conserving. While particle-number non-conserving states still fulfil Eq.~\eqref{eq:stationary} for operators of the form~\eqref{eq:pp_qq}, such that the particle number is conserved \textit{on average}, second-order instabilities (i.e., energy lowering for non-vanishing particle-number variance) are an essential feature of superconducting ground states.

Eventually, scale transformations can also be easily represented by one-body operators~\cite{Bohigas79a}, such that their investigation in the present framework is left for future studies and may be useful in the study of collective modes.

\section{$J$-scheme matrix elements}
\label{app:Jscheme}
The detailed derivation and expressions of the matrix elements entering Eqs.~\eqref{eq:matel} are given in this Appendix. The implementation is kept sufficiently general, i.e., no Hermitian symmetry is assumed, to allow for charge-exchange excitations in future works. The development of this Appendix follows closely from Ref.~\cite{LuJohnson2018}, while also exploiting the theoretical achievements of Refs.~\cite{Chen93a,Chen93b}.

In an angular-momentum coupled basis~\cite{Varshalovich88a} one works with a set of quantum numbers
\begin{equation}
    a=\{j_a m_a l_a n_a\}\,,
\end{equation}
where $j_a$ is the total angular momentum of the $a$ state, $m_a$ its projection on the $z$ axis, $l_a$ the orbital angular momentum and $n_a$ the principal quantum number needed to uniquely defined $a$. In this context, one always wish to work with spherical tensor. In order for the annihilation operators to behave like spherical tensors, one needs to introduce
\begin{equation}
    \tilde{c}_a=(-)^{j_a+m_a}c_{-a}\,.
\end{equation}
General one- and two-body operators are then made up of the following building blocks
\begin{subequations}
\label{eq:build_block}
    \begin{align}
        Q_{JM}(ab)&=\sum_{m_a m_b}(j_a m_a j_b m_b|JM)c_a^\dagger\tilde{c}_b\,,\\
        A_{JM}^\dagger(ab)&=\sum_{m_a m_b}(j_a m_a j_b m_b|JM)c^\dagger_a c^\dagger_b\,,\\
        -\tilde{A}_{JM}(ab)&=\sum_{m_a m_b}(j_a m_a j_b m_b|JM)\tilde{c}_a \tilde{c}_b\,.
    \end{align}
\end{subequations}
An important relation that will be useful in the following is 
\begin{equation}
    Q^\dagger_{JM}(ab)=(-)^{j_a-j_b-M}Q_{J-M}(ba)\,.
\end{equation}
In this way a general one-body operator reads
\begin{equation}
\label{eq:sph_tens}
    O_{LM}=[L]^{-1}\sum_{ab}O_{ab}\,Q_{LM}(ab)
\end{equation}
where
\begin{equation}
    O_{ab}\equiv\braket{a||O_L||b}
\end{equation}
are the reduced matrix elements~\cite{Rose95a}. The Hermitian conjugate of Eq.~\eqref{eq:sph_tens} reads then
\begin{equation}
    O_{LM}^\dagger=[L]^{-1}\sum_{ab}(-)^{j_b-j_a-M}O_{ba}^*\,Q_{L-M}(ab)\,.
\end{equation}
If the condition
\begin{equation}
    O_{ab}=(-)^{j_b-j_a}O_{ba}^*
\end{equation}
is satisfied, the usual properties of Hermitian spherical tensors is recovered
\begin{equation}
    O^\dagger_{LM}=(-)^M O_{L-M}\,.
\end{equation}
In the following, the above property is never used, so to keep the formalism general and applicable also for, e.g., charge-exchange transitions. Let us now give formulas for the commutators between the fundamental building blocks. The derivation in a $J$-coupled formalism can be found in~\cite{Chen93a,Chen93b,LuJohnson2018}. They read
\begin{widetext}
\begin{subequations}
    \begin{align}
    \label{eq:commJ_1}
        [Q_{J_1}(ab),Q_{J_2}(cd)]_J&=[J_1][J_2](-)^{j_a+j_b+j_c+j_d+J}\left[\delta_{bc}\sixj{j_a}{j_b}{J_1}{J_2}{J}{j_d}Q_J(ad)-\delta_{ad}(-)^{J_1+J_2+J}\sixj{j_a}{j_b}{J_1}{J}{J_2}{j_c}Q_J(cb)\right]\,,\\
        [\tilde{A}_{J_1}(ab),Q_{J_2}(cd)]_J&=[J_1][J_2]\left[\delta_{bc}\sixj{j_a}{j_b}{J_1}{J_2}{J}{j_d}\tilde{A}_J(da)-(-)^{J_1+j_a+j_b}\delta_{ac}\sixj{j_a}{j_b}{J_1}{J}{J_2}{j_d}\tilde{A}_J(db)\right]\nonumber\\
        &=[J_1][J_2](1-\mathscr{P}_{abJ_1})\delta_{bc}\sixj{j_a}{j_b}{J_1}{J_2}{J}{j_d}\tilde{A}_J(da)\,,\\
        [A^\dagger_{J_1}(ab),Q_{J_2}(cd)]_J&=[J_1][J_2](-)^{J_2+j_c+j_d}\left[\delta_{bd}\sixj{j_a}{j_b}{J_1}{J_2}{J}{j_c}A^\dagger_J(ca)-(-)^{J_1+j_a+j_b}\delta_{ad}\sixj{j_a}{j_b}{J_1}{J}{J_2}{j_c}A^\dagger_J(cb)\right]\nonumber\\
        &=[J_1][J_2](-)^{J_2+j_c+j_d}(1-\mathscr{P}_{abJ_1})\delta_{bd}\sixj{j_a}{j_b}{J_1}{J_2}{J}{j_c}A^\dagger_J(ca)\,,
    \end{align}
\end{subequations}
where the notation 
\begin{equation}
    \mathscr{P}_{abJ}\equiv(-)^{j_a+j_b+J}P_{ab}
\end{equation}
has been introduced, with $P_{ab}$ the exchange operator between indices $a$ and $b$. The property
\begin{equation}
    [A_{J_a},B_{J_b}]_J=-(-)^{J-J_a-J_b}[B_{J_b},A_{J_a}]_J
\end{equation}
will also be exploited in the following. The Hamiltonian reads, in a $J$-coupled scheme, as 
\begin{equation}
    H=\sum_{ab}h_{ab}[j_a]Q_0(ab)+\frac{1}{4}\sum_{abcd}\zeta_{ab}\zeta_{cd}\sum_J V_J(ab,cd)[J]\left(A^\dagger_J(ab)\otimes\tilde{A}_J(cd)\right)_0\,.
\end{equation}
The main object to be evaluated for the Eqs.~\eqref{eq:matel} is then
\begin{align}
    \sum_M[Y^\dagger_{LM},[H,X_{LM}]]&=[L]^{-2}\sum_{M}\sum_{efgh}(-)^{j_h-j_g+M}Y_{hg}^* X_{ef}[Q_{L-M}(gh),[H,Q_{LM}(ef)]]_{00}\nonumber\\
    &=(-)^L[L]^{-1}\sum_{efgh}(-)^{j_h-j_g}Y_{hg}^* X_{ef}[Q_L^{(2)}(gh),[H,Q_L^{(1)}(ef)]]_0\nonumber\\
    &=-(-)^L[L]^{-1}\sum_{efgh}(-)^{j_h-j_g}Y^*_{hg} X_{ef}[[H,Q_L^{(1)}(ef)],Q^{(2)}_L(gh)]_0\,.
\end{align}
In the following development the aim is to write the matrix elements of the double commutator in the same form as a scalar operator (like the Hamiltonian), such that we want to create an operator of the form
\begin{equation}
    W\equiv\sum_{ab}m_{ab}[j_a]Q_0(ab)+\frac{1}{4}\sum_{abcd}\zeta_{ab}\zeta_{cd}\sum_J \tilde{W}_J(ab,cd)[J]\left(A^\dagger_J(ab)\otimes\tilde{A}_J(cd)\right)_0\,.
\end{equation}
The one-body matrix elements are obtained by the recursive application of Eq.~\eqref{eq:commJ_1}, eventually giving
\begin{align}
    w_{ab}=[j_a]^{-2}\sum_{cd}\left[Y^*_{ca}h_{cd}X_{db}-Y_{ca}^*X_{cd}h_{db}-h_{ac}X_{cd}Y^*_{bd}+h_{ac}Y_{dc}^*X_{db}\right]\,.
\end{align}
The two-body matrix elements are more involved. The double commutator of the two-body part of the Hamiltonian gives.
\begin{equation}
\label{eq:full_W}
    -\frac{1}{4}(-)^L[L]^{-1}\sum_{J}[J]\sum_{abcd}\zeta_{ab}\zeta_{cd}\sum_{efgh}V_J(ab,cd)(-)^{j_h-j_g} Y^*_{hg} X_{ef}\left[\left[\left(A^\dagger_J(ab)\otimes\tilde{A}_J(cd)\right)_0,Q_L^{(1)}(ef)\right],Q^{(2)}_L(gh)\right]_0\,.
\end{equation}
Decomposed into the fundamental blocks from Eqs.~\eqref{eq:build_block}, Eq.~\eqref{eq:full_W} gives the following expression
\begin{align}
    \left[\left[(A^\dagger_J\otimes\tilde{A}_J)_0,Q_L^{(1)}\right]_L,Q^{(2)}_L\right]_0=\sum_{J'}[J'][J]^{-1}[L]^{-1}\bigg\{&(-)^{J+J'+L}&&\left(A^\dagger_J\otimes\left[\left[\tilde{A}_J,Q_L^{(1)}\right]_{J'},Q_L^{(2)}\right]_J\right)_0+\nonumber\\
    & &&\left(\left[A^\dagger_J,Q_L^{(1)}\right]_{J'}\otimes\left[\tilde{A}_J,Q_L^{(2)}\right]_{J'}\right)_0+\nonumber\\
    & &&\left(\left[A^\dagger_J,Q_L^{(2)}\right]_{J'}\otimes\left[\tilde{A}_J,Q_L^{(1)}\right]_{J'}\right)_0+\nonumber\\
    &(-)^{J+J'+L}&&\left(\left[\left[A^\dagger_J,Q_L^{(1)}\right]_{J'},Q_L^{(2)}\right]_J\otimes\tilde{A}_J\right)_0\bigg\}\,.
    \label{eq:4terms}
\end{align}
The four terms (lines) entering Eq.~\eqref{eq:4terms} are computed separately: the first term reads
\begin{align}
    \left[\left[\tilde{A}_J(cd),Q_L^{(1)}(ef)\right]_{J'},Q_L^{(2)}(gh)\right]_J=&[J][J'][L]^2\bigg\{(1-\mathscr{P}_{cdJ})\pi_{fc}^{J'}\delta_{de}\delta_{cg}\sixj{j_c}{j_d}{J}{L}{J'}{j_f}\sixj{j_f}{j_c}{J'}{L}{J}{j_h}\tilde{A}_J(hf)\nonumber\\
    -&(1-\mathscr{P}_{cdJ})(-)^{J'+j_c+j_f}\pi_{fc}^{J'}\delta_{de}\delta_{fg}\sixj{j_c}{j_d}{J}{L}{J'}{j_f}\sixj{j_c}{j_f}{J'}{L}{J}{j_h}\tilde{A}_J(ch)\bigg\}\,.
\end{align}
The second and third terms read
\begin{multline}
    ([A_J^\dagger(ab),Q_L^{(1)}(ef)]_{J'}\otimes[\tilde{A}_J(cd),Q^{(2)}_L(gh)]_{J'})_0=[J]^2[L]^2(-)^{L+j_e+j_f}(1-\mathscr{P}_{abJ})(1-\mathscr{P}_{cdJ})\\
    \times\delta_{bf}\delta_{dg}\sixj{j_a}{j_b}{J}{L}{J'}{j_e}\sixj{j_c}{j_d}{J}{L}{J'}{j_h}(A^\dagger_{J'}(ea)\otimes\tilde{A}_{J'}(hc))_0
\end{multline}
and
\begin{multline}
    ([A_J^\dagger(ab),Q_L^{(2)}(gh)]_{J'}\otimes[\tilde{A}_J(cd),Q^{(1)}_L(ef)]_{J'})_0=[J]^2[L]^2(-)^{L+j_g+j_h}(1-\mathscr{P}_{abJ})(1-\mathscr{P}_{cdJ})\\
    \times\delta_{bh}\delta_{de}\sixj{j_a}{j_b}{J}{L}{J'}{j_g}\sixj{j_c}{j_d}{J}{L}{J'}{j_f}(A^\dagger_{J'}(ga)\otimes\tilde{A}_{J'}(fc))_0
\end{multline}
respectively. Eventually, the last term of Eq.~\eqref{eq:4terms} reads
\begin{align}
    \left[\left[A^\dagger_J(ab),Q_L^{(1)}(ef)\right]_{J'},Q_L^{(2)}(gh)\right]_J=&[J][J'][L]^2(-)^{j_e+j_f+j_g+j_h}\bigg\{(1-\mathscr{P}_{abJ})\pi_{ea}^{J'}\delta_{bf}\delta_{ah}\sixj{j_a}{j_b}{J}{L}{J'}{j_e}\sixj{j_e}{j_a}{J'}{L}{J}{j_g}A_J^\dagger(ge)\nonumber\\
    -&(1-\mathscr{P}_{abJ})(-)^{J'+j_a+j_e}\pi_{ea}^{J'}\delta_{bf}\delta_{eh}\sixj{j_a}{j_b}{J}{L}{J'}{j_e}\sixj{j_a}{j_e}{J'}{L}{J}{j_g}A^\dagger_J(ga)\bigg\}\,.
\end{align}
The above expressions are inserted into Eq.~\eqref{eq:full_W} in order to deduce the two-body matrix elements of $W$, which are decomposed in the following way
\begin{equation}
\label{eq:six_terms}
    \tilde{W}_J(ab,cd)=\sum_{i=1}^6 \tilde{W}_i(abcd,J)\,,
\end{equation}
where the six terms entering Eq.~\eqref{eq:six_terms} are derived as follows
\begin{subequations}
\begin{align}
    W_1(abcd,J)&=-\sum_{\substack{ef\\J'}}(-)^{J+J'}[J']^2\pi_{de}^{J'}\zeta^{-1}_{cd}\zeta_{ef}(-)^{j_c-j_e}Y^*_{ce}X_{fd}\sixj{j_e}{j_f}{J}{L}{J'}{j_d}\sixj{j_d}{j_e}{J'}{L}{J}{j_c}
    V_J(ab,ef)\,,\\
    W_2(abcd,J)&=-\sum_{\substack{ef\\J'}}[J']^2\pi_{cf}^{J'}\zeta_{cd}^{-1}\zeta_{ce}Y_{df}^* X_{ef}\sixj{j_c}{j_e}{J}{L}{J'}{j_f}\sixj{j_c}{j_f}{J'}{L}{J}{j_d}
    V_J(ab,ce)\,,\\
    W_3(abcd,J)&=\sum_{\substack{ef\\J'}}[J']^2\zeta_{ab}^{-1}\zeta_{cd}^{-1}\zeta_{be}\zeta_{df}(-)^{j_c-j_f}Y_{cf}^* (-)^{j_a-j_e} X_{ae}\sixj{j_b}{j_e}{J'}{L}{J}{j_a}\sixj{j_d}{j_f}{J'}{L}{J}{j_c}
    V_{J'}(be,df)\,,\\
    W_4(abcd,J)&=\sum_{\substack{ef\\J'}}[J']^2\zeta_{ab}^{-1}\zeta_{cd}^{-1}\zeta_{bf}\zeta_{de}Y^*_{fa} X_{ec}\sixj{j_b}{j_f}{J'}{L}{J}{j_a}\sixj{j_d}{j_e}{J'}{L}{J}{j_c}
    V_{J'}(bf,de)\,,\\
    W_5(abcd,J)&=-\sum_{\substack{ef\\ J'}}(-)^{J+J'}[J']^2\zeta_{ab}^{-1}\zeta_{ef}\pi_{be}^{J'}Y^*_{ea} (-)^{j_b-j_f}X_{bf}\sixj{j_e}{j_f}{J}{L}{J'}{j_b}\sixj{j_b}{j_e}{J'}{L}{J}{a}
    V_J(ef,cd)\,,\\
    W_6(abcd,J)&=-\sum_{\substack{ef\\J'}}[J']^2\pi_{af}^{J'}\zeta_{ab}^{-1}\zeta_{ae}(-)^{j_f-j_b}Y_{fb}^*(-)^{j_f-j_e}X_{fe}\sixj{j_a}{j_e}{J}{L}{J'}{j_f}\sixj{j_a}{j_f}{J'}{L}{J}{j_b}
    V_J(ae,cd)\,.
\end{align}
\end{subequations}
The above derivation has exploited the anti-symmetrization of the matrix elements of $H$~\cite{Suhonen07a}. In order for the two-body matrix elements of $W$ to be anti-symmetrized as well, one eventually writes
\begin{subequations}
    \begin{align}
        \tilde{W}_1(abcd,J)&\equiv(1-\mathscr{P}_{cdJ})W_1(abcd,J)\,,\\
        \tilde{W}_2(abcd,J)&\equiv(1-\mathscr{P}_{cdJ})W_2(abcd,J)\,,\\
        \tilde{W}_3(abcd,J)&\equiv(1-\mathscr{P}_{abJ})(1-\mathscr{P}_{cdJ})W_3(abcd,J)\,,\\
        \tilde{W}_4(abcd,J)&\equiv(1-\mathscr{P}_{abJ})(1-\mathscr{P}_{cdJ})W_4(abcd,J)\,,\\
        \tilde{W}_5(abcd,J)&\equiv(1-\mathscr{P}_{abJ})W_5(abcd,J)\,,\\
        \tilde{W}_6(abcd,J)&\equiv(1-\mathscr{P}_{abJ})W_6(abcd,J)\,,
    \end{align}
\end{subequations}
which complete the derivation of the six addends entering Eq.~\eqref{eq:six_terms}. 
Eventually, one notices that the following properties are found for $X$ and $Y$ being Hermitian spherical tensors satisfying Eq.~\eqref{eq:sph_tens}
\begin{subequations}
    \begin{align}
        W_4(abcd,J)&=P_{ac}P_{bd}W_3^*(abcd,J)\,,\\
        W_5(abcd,J)&=P_{ac}P_{bd}W_1^*(abcd,J)\,,\\
        W_6(abcd,J)&=P_{ac}P_{bd}W_2^*(abcd,J)\,.
    \end{align}
\end{subequations}
\end{widetext}
In such case, the derived expression are fully equivalent to those provided by Ref.~\cite{LuJohnson2018}. In this respect, thus, this work generalizes the derivation from Ref.~\cite{LuJohnson2018} to the case of $X$ and $Y$ being different operators and, possibly, non-Hermitian.

\bibliography{biblio}

\end{document}